\documentclass[nofootinbib,floats,aps,superscriptaddress,
               preprint]{revtex4} 
\usepackage{graphicx}
\newcommand{\bea}{\begin{eqnarray}}
\newcommand{\eea}{\end{eqnarray}}
\newcommand{\beq}{\begin{equation}}
\newcommand{\eeq}{\end{equation}}
\newcommand{\bay}{\begin{array}}
\newcommand{\eay}{\end{array}}

\def\umax{u_{\rm max}} 

\def\sign{{\rm sign}} 

\def\gsim{\ \rlap{\raise 3pt \hbox{$>$}}{\lower 3pt \hbox{$\sim$}}\ }
\def\lsim{\ \rlap{\raise 3pt \hbox{$<$}}{\lower 3pt \hbox{$\sim$}}\ }

\begin{document}  
 
\preprint{\vbox{\hbox{WIS/42/02-OCT-DPP~~}\hbox{hep-ph/0210053}
   \hbox{October, 2002}}}
  
\vspace*{1cm}  
  
\title{\boldmath Realistic construction of split fermion models}  
  
\author{Yuval Grossman}  
\affiliation{Department of Physics,  
Technion--Israel Institute of Technology,\\ 
Technion City, 32000 Haifa, Israel\vspace{6pt}}
  
\author{Gilad Perez}
\affiliation{Department of Particle Physics, Weizmann Institute of Science\\
 Rehovot 76100, Israel
	\\[20pt] $\phantom{}$ }

\begin{abstract} \vspace*{8pt}  
The Standard Model flavor structure can be explained in theories where
the fermions are localized on different points in a compact extra
dimension. We show that models with two bulk scalars compactified on
an orbifold can produce such separations in a natural way. We study
the shapes and overlaps of the fermion wave functions.  
We show that, generically,
realistic models of Gaussian overlaps are unnatural since they require
very large Yukawa couplings between the fermions and the bulk scalars.
We give an example of a five dimensional two scalar model that
accounts naturally for
the observed quark masses, mixing angles and CP violation.
\end{abstract}
  
\maketitle

\section{Introduction}
The Standard Model (SM) flavor puzzle is to understand the origin of
the smallness and hierarchy of most of the fermion masses and mixing
angles. The likely solution of this puzzle requires the existence of a
fundamental theory that generates the observed patterns in a
natural way, namely without small dimensionless parameters.
One such framework is due to Arkani-Hamed and Schmaltz (AS)
\cite{AS}. The idea is to separate the various SM fermion fields
inside compact extra dimensions. In that case the four dimensional
Yukawa couplings between two fermions are suppressed by the overlap of
their zero mode wave functions. Once the widths of the wave functions are
smaller than their separation in the extra dimensions, small and
hierarchical Yukawa couplings in four dimensions are induced.

Various scenarios based on this idea had been constructed. In some
models, only the fermion wave functions have non trivial shapes in the
extra dimension \cite{AS,MS,GGH,KT,ASmodels,Mats,GP}. 
In other cases, also the Higss
field is used to generate the flavor structure \cite{KT,GaussLocH,LocH}.  
A specific model in five dimensions (5d) that generates the observed
fermion masses and mixing angles was found by Mirabelli and Schmaltz
(MS) \cite{MS}. This model, however, cannot accommodate the observed
CP violation in the kaon and $B$ systems \cite{Branco}.  Experimental
tests of the split fermion idea were studied in \cite{AS,AGS}.

While the AS framework is very attractive there is no complete
realization of it. In \cite{AS} a model with one infinite extra
dimension, $x_5$, was considered where domain wall fermions were used
to localize the fermions and to separate them.  The basic idea is that
there is a scalar, $\Phi$, which develops an $x_5$ dependent vacuum
expectation value (vev),
$h(x_5)$. Then, the fermions have $x_5$ dependent effective masses
$m_{eff}=\lambda h(x_5)-M$, where $M$ is a bare mass term and
$\lambda$ is the Yukawa coupling between the fermion and $\Phi$. The
minimum of the right handed (RH) zero mode and the maximum of the left
handed (LH) zero modes is where the effective mass vanishes, namely at
$x_5^m$ such that $\lambda h(x_5^m)=M$.  The LH chirality state, with
the maximum at $x_5^m$ is localized around $x_5^m$. The RH mode,
however, is localized at infinity, namely it is non-normalizable and
it drops out of the spectrum. The separation between the LH modes is
achieved by introducing a different $M$ for the different fields.

The generalization of this original realization of the split fermions
idea to the realistic case of a compact dimension is not trivial. The
problem is that in a compact space both LH and RH zero modes are
normalizable, and thus the model is not really chiral. 
Another problem, which is related to the first one, is that in a compact space
the domain wall configuration is not stable.

Some ideas toward solving these problems had been proposed by Georgi,
Grant and Hailu (GGH) \cite{GGH} and by Kaplan and Tait (KT)
\cite{KT}.  They considered 5d scenarios with one extra dimension
compactified on an $S_1/Z_2$ orbifold.  The $Z_2$ orbifold symmetry
projects out one chiral zero mode, making this a chiral theory in four
dimensions (4d). A scalar, $\Phi$, which is odd under the $Z_2$
symmetry is also introduced. Since $\Phi$ is odd, its vev must vanish
on the orbifold fixed points. When $\Phi$ develops a non zero vev in
the bulk, an $x_5$ dependent vev is generated.  Fermion mass terms of
the from $\Phi \bar \Psi_R \Psi_L$ are allowed while bare mass terms,
$\bar\Psi_R \Psi_L$, are forbidden by the $Z_2$ symmetry.  Therefore,
the zero modes are localized around one of the boundaries since
these are the only places where their effective masses vanish. The sign of the
Yukawa coupling between the fermion and the scalar determines on which
boundary the fermion is localized.  The idea of split fermions where
each fermion is stuck at a different point in the bulk cannot be
achieved. The zero modes can be localized only around one of the orbifold
fixed points. An additional flavor structure in the Yukawa couplings
is needed in order to naturally generate the observed fermion
parameters in such two location models \cite{KT}.  More complicated
models, using large non renormalizable\footnote{More precisely, we
refer here to 5d terms which after dimensional reduction produce non
renormalizable operators in the 4d effective theory.}  terms
\cite{GGH} or boundary terms \cite{KT} can localize fermions in the
bulk.

The problems of this setup can be traced down to two points.
The first is that the scalar vev vanishes only on the
boundaries. Thus, the fermions cannot be localized within the bulk. The second
is that mass terms are absent. In the original AS model the
scalar was used to localize the fermion zero modes, and different
constant mass terms were used to split them. In the simple orbifold
models, while the scalar can be used to localize the fermions, there is
no simple mechanism that splits them.

We are therefore interested in improving upon the models of
\cite{GGH,KT}.  We propose to use two scalar fields that couple to
the fermions.  This rather minor modification of the GGH and KT
scenarios can produce localization in the bulk and at the same time does
not affect the appealing features of these models.  The advantage that
two scalar models have over one scalar models is that the effective
mass can vanish in the bulk.  Intuitively the picture is as
follows. For one scalar the sign of the Yukawa coupling determines the
boundary where the fermion is localized \cite{GGH}. Once a second
scalar with opposite sign Yukawa coupling is introduced, the picture
is more complicated.  The second scalar tends to localize the fermion
on the other boundary.  In some cases one scalar is dominant and
localization is at the boundary determined by the sign of its Yukawa
coupling and the second scalar only modifies the shape of the wave
function. In other cases, however, the tension between the two scalar
results in a compromise: A configuration where the fermion is
localized in the bulk.

In section II we introduce the two scalar model and investigate its
scalar sector.  In section III we study the shapes of the fermion
wave functions in various limits of the model and their implications
for model building.  In section IV we present a two scalar model
which accounts naturally for the quark masses, mixing angles and CP
violation.  We conclude in section V.

\section{Two scalar model}
We now move to study the features of the scalar sector of our model.
We explain how it can generate localization and separation of fermion
wave functions in the extra dimension.

The space-time of our model is described by ordinary 4d space-time and
an additional space dimension compactified on an $S_1/Z_2$
orbifold. The physical region is defined as $0 \le x_5 \le L$ and
$x_5=0,L$ are the orbifold fixed points. The model includes two scalar
fields, $\Phi_i$ ($i=1,2$), and a fermion field, $\Psi$.  The scalars
and $\Psi_R$ are odd while $\Psi_L$ is even under the $Z_2$ orbifold
symmetry.  For simplicity, we assume that there is no interaction
between the two scalars.  We further consider the case where the two
scalars develop $x_5$ dependent vevs. The conditions for this case to
be realized can be found by a straightforward generalization of the
single scalar case (see e.g. \cite{GP}). Here we only remark that
having an $x_5$ dependent vev is a generic situation.  Then, the
action is
\beq 
{\cal S}=\int dx_M
\left\{\bar\Psi\left(i\gamma^M\partial_M
+\tilde f_1\Phi_1+\tilde f_2\Phi_2\right)\Psi
+\sum_i\left[{1\over2}
\partial^M\Phi_i\partial_M\Phi_i
-\tilde\lambda_i\bigl(\Phi_i^2-\tilde v_i^2\bigr)^2\right] \right\}
\,. \label{L5}
\eeq
where $\tilde f_i$ and $\tilde\lambda_i$ are real with $\tilde
\lambda_i>0$.  (Our notation is such that capital letters run over
the five dimensions while Greek letters over the four dimensions.)

It is convenient to use dimensionless parameters. We therefore use
rescaled fields
\beq \label{scale1}
\Psi\to {\psi\over \sqrt{L}}\,, \qquad
\Phi \to {\varphi \over \tilde v}\,,
\eeq
and introduce the following dimensionless quantities
\beq \label{scale2}
u\equiv {x_5 \over L}\,, \qquad
a_i\equiv \sqrt{2\tilde\lambda_i}\, \tilde v_i L\,, \qquad
f_i\equiv {\tilde f_i\over \sqrt{2 \tilde \lambda}}\,, \qquad
X\equiv -{a_2 f_2 \over a_1 f_1}\,.
\eeq
Note that $0 \le u \le 1$. We  
order the two scalars such that $a_1 < a_2$ and
further define $f\equiv f_1$ and $a\equiv a_1$.
Then, the action can be written as
\bea{\cal S}&=&
\int dx_\mu\int_0^1 du \left\{
\left[\bar\psi\left(i\gamma^\mu\partial_\mu-
{1\over L}\gamma_5\partial_5-{f a\over L}\left(
\varphi_1-X\varphi_2\right)\right)\psi\right]\right. 
\nonumber \\ 
&+&\left.
\sum_i {\tilde v^2_i\over L}\left[
L^2 {1\over2}\partial^\mu\varphi_i\partial_\mu\varphi_i
-{1\over2}\partial_5\varphi_i\partial_5\varphi_i
-{a^2_i\over2}\left(\varphi^2_i-1\right)^2\right]\right\}
\,, \label{L5D2S}
\eea
where $\partial_5\equiv {\partial\over \partial u}$. 

We move to investigate the functions
$h_i(u)\equiv\langle\varphi_i\rangle(u)$.  For one scalar they were
investigated in detail in \cite{GGH,GP}. Since we assume that there
is no interaction between the scalars we can use the one scalar model
results for each of the scalars. Neglecting the Yukawa
interactions, and working in the limit where \cite{GP}
\beq
a_i \gg 1,
\eeq
we have
\beq\label{hi}
h_i(u)=\tanh\left[a_i u\right] \tanh\left[a_i (1-u)\right] \,.
\eeq
Since $h_i(u)$ is symmetric under $u\to 1-u$ we consider
only the $0 \le u \le 1/2$ range.
We will make use of the following approximations
\beq
h_i(u)\approx\cases{1 & for~ $u \gg 1/a_i\,,$\cr
a_i u & for~ $u \ll 1/a_i\,$.}
\label{hap}
\eeq 
Of particular interest is the following function
\beq\label{gu}
g(u)= h_1(u) - X h_2(u)\,.
\eeq
Given the values of $a_i$ and recalling that $a_1<a_2$, 
the sign of $g(u)$ is determined by $X$ as follows:
\begin{itemize}
\item[(i)] $X>1$
\bea\label{ng}
\sign[g(u)]=-1\,.
\eea
\item[(ii)] $X<a_1/a_2$
\bea\label{pg}
\sign[g(u)]=1\,.
\eea
\item[(iii)] $1>X>a_1/a_2$  
\bea\label{alter}
\sign\left[g\left(u\ll {1\over a_2}\right)\right]=-1\,,\qquad 
\sign\left[g\left(u\approx {1\over 2}\right)\right]=1
\,.
\eea
\end{itemize}
Eq. (\ref{alter}) teaches us that when condition (iii) is satisfied
$g(u)$ vanishes not only on the boundaries. Namely, there exists a
point $0<u_{\rm max}<1/2$ such that $g(u_{\rm max})=0$.  (Note that
since $h_i(u)$ is symmetric under $u\to 1-u$ if $g(u)$ crosses zero in
the bulk, it crosses it twice, at $u_{\rm max}$ and $1-\umax$.)

While we cannot solve analytically for
$u_{\rm max}$ in the general case, we can do it in what
we denote as the constant mass approximation
\beq\label{conM}
a_1/a_2 \ll 1
\,.
\eeq 
In this approximation $u_{\rm max}$ is given by
\beq \label{umaxre}
u_{\rm max} \approx {{\rm arctanh(X)}\over a} = {X \over a} +O(X^3)\,.
\eeq

Figs. ({\bf 1b}) and ({\bf 1c}) 
show the function $g(u)$ (the thin black curves) in the cases described by
eq. (\ref{alter}) for typical values of parameters.  The figures
demonstrate the main point of this section: The two scalar model yields
an effective scalar profile which has several zeros. In particular,
one of them is not an orbifold fixed point. Thus, we can construct
models in which the fermions are localized at various points in the
fifth dimension and not only at the orbifold fixed points.

\section{The fermion zero mode}
In the following we analyze the functional behavior of the fermion
zero mode wave function and, once more fermions are introduced, the
overlaps between their zero mode wave functions in the fifth
dimension.  Though we do it in the context of a two scalar model we
shall see that some of our conclusions are rather generic.  They apply
to other realizations of the split fermions scenario.

Following the standard treatment (see e.g. \cite{AS}) we find the
differential equation for $y(u)$, the fifth dimension wave function of
$\Psi_L\,$:
\beq
{1 \over y(u)}{\partial y(u) \over \partial u}=-fa \,g(u)
\label{EL3}
\,,
\eeq
where $g(u)$ is defined in (\ref{gu}).  The solution is given by
\beq \label{solkx}
y(u)=N\exp\left[-fa\int_0^{u}\, g(w)\, dw \right]\,,
\label{psi5}
\eeq
where $N$ is the normalization factor.  The local maxima of $y(u)$ are
at points in which $g(u)$ vanishes.  Which of these maxima is the
global maximum depends on the model parameters, and can be at any of
these points.

\subsection{A single scalar model}\label{ss}
We begin our analysis by  setting $X$ to zero in eq. (\ref{gu})
and returning to the single scalar case \cite{AS,KT}.
Then, the fermion zero mode is given by
\beq
y(u) \simeq 
N\; \exp\left\{-f\,\ln\left[\cosh(au)\right] \right\}
{\rm ~~~for~~~} u\lsim 1/2
\,.\eeq
We shall analyze $y(u)$ in two limits, large and small $f$.  We show that
these are related to the two scenarios studied by AS \cite{AS}
and KT \cite{KT} respectively.  We assume that $f$ is positive and
thus the maximum of $y(u)$ is at $u=0$. Then, we have
\bea
y(u) &\approx& 
N \; \exp\left[-{f\,a^2\,  \over 2}\,u^2\right]
{\rm ~~~for~~~} u\ll 1/a\,;\label{yll} \\
y(u) &\approx& 
N  \; \exp\left[-f(a\,u-\ln2)\right]
{\rm ~~~for~~~} 1/a\lsim u\lsim 1/2\,.
\label{y1}
\eea
When eq. (\ref{yll}) is applicable, the fermion wave function
is approximately a Gaussian with a width
\bea \label{GG}
\Gamma\sim {1\over a \sqrt f}
\,.\eea
When eq. (\ref{y1}) is applicable, the fermion wave function
is approximately an exponential with a width
\bea \label{Ge}
\Gamma \sim {1\over a f}
\,.\eea
Substituting $u= 1/a$ in (\ref{yll}) we learn that when 
$f\gg1$ the exponential part is negligible
and then $y(u)$ is of a  Gaussian shape.
Similarly, substituting $u= 1/a$ in (\ref{y1}) we learn that for
$f\lsim1$ the Gaussian part is negligible.

\subsection{A two scalar model}\label{ts}

We now move to the two scalar scenario.
While the integral in (\ref{solkx}) can be solved analytically, 
the result is not very illuminating.
We therefore consider $y(u)$ in various limits related to 
the three cases of eqs. (\ref{ng}), (\ref{pg}) and (\ref{alter}). 

When $X>1$ or $X<a_1/a_2$ the function $g(u)$ vanishes only on the
fixed points and $y(u)$ is localized at either $u=0$ or $u=1$.  This
case is similar to the single scalar model (with a zero mass)
discussed above. The corresponding fermion wave function will be of
the form of either Gaussian or exponent as shown in eqs. (\ref{yll})
or (\ref{y1}).  In the limit of large $a_i$ and with $f$ of order
unity it reproduces the KT exponential model.

When $1>X>a_1/a_2$ the fermions can be localized in the bulk.
It is useful to discuss this case in the constant mass approximation
[see eq. (\ref{conM})].
In this approximation the second term in $g(u)$  
can be treated as a constant and then 
\beq\label{CM}
y(u) \simeq 
N\; \exp\left\{-f\left[\ln\left[\cosh(au)\right]-X\right] \right\}
{\rm ~~~for~~~} u\lsim 1/2
\,.\eeq
Substituting $X=M L/af$
in (\ref{CM}), with $M$ being the 5d fermion mass,   
the model reproduces the constant mass models of \cite{AS,MS}.
Thus all our following analysis and conclusions apply to these models
as well.

We like to approximate $y(u)$ near its maxima.
The situation in this case is more involved since $y(u)$ has two
maxima and which is the dominant one depends on the interplay between 
the values of  $X$, $a$ and $f$. 
Inspecting Eq. (\ref{solkx}) we see that this is determined
by the relative size of $I_p$ and $I_n$ defined as
\beq \label{defipin}
I_p=\int_0^{\umax} g(u) du, \qquad
I_n=\int_{\umax}^{1-\umax} g(u) du\,.
\eeq
Once $I_p>I_n$ the global maximum is at one of the fixed points, while in the
opposite case it is at $\umax$ or $1-\umax$. 
We define $X_b$ to be the transition point, namely, the point where
$I_p=I_n$.
Using the constant mass approximation [eq. (\ref{CM})] and the linear
approximation for $h_1$ [eq. (\ref{hap})] we find the following
estimate for the transition point: 
\beq\label{twoM}
X_b\sim{2a-4\over 2a-3}\,.
\eeq

We therefore consider three cases:
\begin{itemize}
\item $X > X_b$. 
The global maximum is at one of the orbifold fixed points. This case
is similar to the case of a function with only one maximum and does not
lead to a new structure.
\item $X < X_b$.
The dominant maximum is in the bulk, $0<\umax<1$,
and $y(u)$ is approximately given by
\beq
y(u) \approx 
N\; \exp\left[-{f\,a^2\,\left(1-X^2\right) \over 2}\,
\left(u-\umax\right)^2\right]
{\rm ~~~for~~~}|u-\umax|\ll 1/a\,.\label{GG2a}
\eeq
This is a Gaussian with a maximum at $\umax$ and a width
\bea \label{GG2}
\Gamma={1\over a \sqrt{ f(1-X^2)}}\,.
\eea
For $X\ll1$ the width is independent of $X$ [as in eq. (\ref{GG})], and
the maximum location is
linear in $X$. This is the case
discussed in \cite{AS,MS} (with the substitution $X=ML/af$). 
The generic situation, however, is that the peak of the wave function
is not linear with $X$ and that its width 
grows with the distance of the peak from the origin. 
\item $X\sim X_b$. 
The two maxima are significant. This case cannot be described analytically
but it is a combination of the above two cases.
We remark that this transition region between the other two cases
considered above
becomes sharper as the value of 
$f$ increases. Since we are working with moderate values
of $f$ it does not require fine tuning of $X$ to have a situation
where both peaks are important.
\end{itemize}

In figures {\bf (1a)} -{\bf (1c)} we plot the effective two scalar vev,
$g(u)$, and the corresponding fermion wave function for these three cases.

\subsection{Wave functions overlaps} \label{Gproblem}
We extend the model by including the SM fermion fields and one Higgs
field. After compactification, the zero modes of these fields consist
of the 4d SM fields. In particular, we are interested in how the 4d
Yukawa interactions are generated form the 5d theory. Consider, for
example, the 5d Yukawa interaction between a quark doublet ($Q$), a
down type quark singlet ($D$) and the Higgs field ($H$)
\beq
{\cal L} = {Y_5 \over \sqrt{M_*}} \,\bar Q H D + h.c.,
\eeq
where $M_*$ is the natural scale of the 5d theory. After
compactification the 4d Yukawa
interaction is given by
\beq
{\cal L} = Y_5 K \bar d_L h d_R + h.c.,
\eeq
where lower case letters represent 4d zero mode fields.
The overlap $K$ is defined as
\beq
K \equiv \int_0^1 y_D(u) y_Q(u) du\,,
\eeq
where $y_D$ ($y_Q$) is the fifth dimension wave function of the
singlet (doublet) field. We assumed here that the Higgs vev is flat in
$u$. While this is not generally the case, we assume that its shape is
not far from flat, and thus approximating it by a flat profile is
reasonable. Assuming $Y_5 \sim O(1)$ the flavor structure arises from
the various overlap integrals \cite{AS}.

Suppose that we consider 
a realistic 5d model which accounts for the quark
masses and mixing angles by Gaussian like suppressions.  (Such
overlaps were assumed, for example, in constructing the MS model
\cite{MS} and in many other papers see
e.g. \cite{ASmodels,GaussLocH}.)  The model is specified by the Yukawa
couplings $f^i$ and effective masses $X^i$ of the nine SM quark
fields.
We ask what are the ranges of the 5d flavor parameters in order for
the above model to yield the correct masses and CKM matrix elements.

In order to obtain Gaussian suppressed overlaps and to reproduce the
MS model several conditions are required. First, the significant
overlap between any two fermion wave function has to be in the region
where both wave functions are, to a good approximation, Gaussian.
Second, their maxima have to scale linearly with $X$.  The first
condition is satisfied in the large $f$ limit.  The second condition
requires small $X$.

To make our discussion concrete we assume that the most distant
fermions have $X\lsim0.3$ which implies $\umax\sim 0.3/a$.  Using
Eq. (\ref{GG2a}) we find that the overlap between the Gaussian wave
functions, assuming universal couplings $f$, is given by
$\exp(-fX^2)$.  In \cite{MS} it was found that a viable model of quark
masses and mixing requires a separation of roughly 18 standard
deviations between the two most distant quarks.  Then, for $X\lsim0.3$
we find
\bea \label{largef}
f\gsim 2\left({18 \over 0.3}\right)^2 \sim 10^4
\,.
\eea
Assuming perturbativity, this implies that $\tilde \lambda_i$
[introduced in (\ref{L5})] must be much smaller than their natural
values \cite{CML}.  Note that to a large extent, condition (\ref{largef}) is
independent of the fundamental theory parameters such as $L$, $a$ and
$M_*$.  Thus, it is generic that models with Gaussian
overlaps require very large couplings $f$.

While the model of pure Gaussian wave functions seems unnatural, 
using the analysis of subsection \ref{ts}
it is possible to construct models which account naturally for the
quark flavor parameters.  
This is demonstrated in the next section.

\section{Toward a realistic natural model}
As already mentioned, several models that produce the observed fermion
masses and mixing angles had been constructed.
Some of them use Higgs wave functions that are not flat to generate
the flavor structure.  We do not discuss these models here. Instead we
discuss models that assume, similar to our model, 
that the Higgs vev do not play a
significant role in generating the observed flavor patterns. In the
Appendix we discuss the GGH model \cite{GGH} since our model is not
directly related to it. Below we describe the KT model \cite{KT} and
our improved version of it.

The KT setup is as follows.  LH fermions are localized on one of the
orbifold fixed points, and RH fermions on the other one. The different fermion
wave functions have different widths such that the lighter the fermion
is, its LH and RH wave functions are narrower.  This induces
hierarchical structures for the corresponding mass matrices. Yet, in
order to produce the observed quark masses and mixing angles, this
structure is not enough and the 5d Yukawa couplings must also have
some flavor structure \cite{KT}.

One reason for that, for example, is related to the large top mass.
In order to generate the $O(1)$ 4d Yukawa coupling, the overlap
between $y^{Q_3}$ and $y^{u_3}$ has to be large. Since these wave
functions are localized on different orbifold fixed points, it implies
that at least one of them has to be very broad. Both possibilities, however
are problematic. When $y^{Q_3}$ is broad, a large mass for the
$b$ quark is generated, in contrast to observation. When $y^{u_3}$ is
broad the overlaps between it and the other doublets are large.  Then,
large CKM mixing angles are generated, in disagreement with the data.
Similar problems are encountered when one tries to account for the
flavor parameters of the other generations.

The KT model is significantly improved once some of the fields are
localized in the bulk.  For example, the problem with the third
generation is solved once $y^{Q_3}$ and $y^{u_3}$ are
localized near each other far from any of the orbifold fixed points.
Then, their overlap is large and large $m_t$ is produced. At the same
time, the overlaps of $y^{u_3}$ with the other quark doublet wave functions are
small, thus producing small CKM mixing angles. The $b$ quark mass is
produced by constructing $y^{d_3}$ to have a double peak shape, such
that it has moderate overlap with $y^{Q_3}$.
 
Based on these ideas we construct an example of a configuration which
roughly reproduces the correct quark masses and mixing angles. We
choose the following parameters:
\bea\label{MSbar}
a_1=4\,,&&\qquad a_2=12\,;\nonumber\\
f_1^{Q_{1,2,3}}=30,4,40\,,&&\qquad  
  X^{Q_{1,2,3}}=-4,-1.9,0.73\,;\nonumber\\
f_1^{u_{1,2,3}}=-1,1,28\,,&&\qquad  
  X^{u_{1,2,3}}=-2.3,2.3,0.78\,;\nonumber\\
f_1^{d_{1,2,3}}=-1,-1,21\,,&&\qquad  
  X^{d_{1,2,3}}=-2,-1.4,0.87\,.
\eea
The resultant shapes of the various wave functions are shown in
figs. {\bf (2a)} and  {\bf (2b)} for the up and down type
quarks respectively.  The mass matrices of the up type and down type
quarks are given, up to order one coefficients, by (in MeV units)
\bea
\label{Mup}
M^u\sim\pmatrix{
       1.9&3.1\cdot10^2&6.4\cdot10^2\cr
       4.7&7.2\cdot10^2 &4.2\cdot10^3\cr
       1.7\cdot10^2&3.8\cdot10^3&1.6\cdot10^5}
\quad
M^d\sim\pmatrix{
       5.3&4.0\cdot10^1&8.8\cr
       1.3\cdot10^1&9.3\cdot10^1 &5.3\cdot10^1\cr
       2.4\cdot10^2&7.2\cdot10^2&3.4\cdot10^3}
\,.
\eea 
Their eigenvalues are given by
\bea &&m_u\sim 1.9\,{\rm MeV} \,,\qquad m_c\sim 7.2\cdot10^2\,
{\rm MeV} \,,\qquad
m_t\sim 1.6\cdot10^5\,{\rm MeV} \,,\nonumber \\
&& m_d\sim 5.3\,{\rm MeV} \,,\qquad m_s\sim 9.3\cdot10^1\,{\rm MeV} 
\,,\qquad m_b\sim 3.4\cdot10^3\,{\rm MeV}  \,.\eea 
which agree with the known quark masses, calculated at the $M_Z$ scale 
\cite{Mass,PDG}.  The rotation
matrices from the left that diagonalize the mass matrices are given by
\bea
\label{VupL}
V_L^u\sim\pmatrix{
       1&2\lambda&0.7\cdot\lambda^3\cr
       2\lambda&1&0.5\cdot\lambda^2\cr
       0.4\lambda^3&0.6\cdot\lambda^2&1} \quad
V_L^d\sim\pmatrix{
       1&2\lambda&0.4\cdot\lambda^3\cr
       2\lambda&1&0.4\cdot\lambda^2\cr
       0.5\cdot\lambda^3&0.4\cdot\lambda^2&1}\,,
\eea
where $\lambda=0.22$.  The absolute values of the elements of the CKM
matrix, $V_{CKM}= (V_L^u)^\dagger V_L^d$, are given to leading order in
powers of $\lambda$ by
\bea \label{CKM} 
\left|V_{CKM}\right|\sim \pmatrix{
1&\lambda&\lambda^3\cr \lambda&1&\lambda^2\cr
\lambda^3&\lambda^2&1}\,.  
\eea 
The values of the quark mixing angles agree with the experimental data
\cite{PDG}.  Assuming arbitrary phases we find for the Jarlskog
measure of CP violation \cite{Jcp}
\beq
J \sim 10^{-5}\,,
\eeq
in agreement with the value obtained from global fit to the available
data \cite{PDG}.  We conclude that the above model also reproduces the
observed size of CP violation. (The same is true also in the original
KT model \cite{KT}.) This is in contrast to the MS model, where the
observed order of magnitude of $J$ cannot be accommodated
\cite{Branco}.  The reason is that in the MS model there are very
small entries due to the strong Gaussian suppressions. Here this is not
the case.  The matrices $M^u$ and $M^d$ do not contain such tiny
entries.  Actually, the generic situation is that split fermion models
in five dimensions that produce the correct masses and mixing angles
also account for the observed CP violation.

The example above is only a demonstration that the observed fermion
masses and mixing angles can be generated. The apparent small
deviations from the experimental values are due to the unknown Yukawa
couplings. The point we emphasize is that the flavor hierarchy arises
only from the overlap between the different wave functions. The
unknown $O(1)$ Yukawa couplings do not carry any hierarchical 
flavor structure.

\section{Discussion and Conclusion}

There are several related issues that we did not study and seem to us
worth further investigation.  We did not discuss the lepton sector. 
It is easy
to generate a model for the charged lepton masses. The issue of
neutrino masses is more complicated and may require additional
ingredients. We also did not investigate the Higgs vev shape. While we
assumed that it is flat, generally this is not the case. It seems to
us that it can be flat enough in order not to upset our general
conclusions. However, it is interesting to fully understand the vev
shape and its implications.

Our model can be extended, for example, by adding more scalars.  Then,
the number of the fermion wave functions maxima can be more then two,
up to the number of scalars. One of the maxima is an orbifold fixed
point, and the others are in the bulk. Having more than one maxima in
the bulk can be used to obtain a richer structure.
However, this extra flexibility is not needed in order to generate the
observed quark masses and mixing angles.

The framework we considered cannot be directly related to the
fundamental theory of nature. It must be a low energy limit of a more
fundamental theory that should also explain the compactification and
stabilization of the orbifold setup.  Therefore, in general, non
renormalizable terms and terms localized on the orbifold fixed points
can be present in the effective theory. Since such terms
modify the shape of the fermion wave functions they were used in
\cite{GGH} and \cite{KT} to generate the needed structure. These terms, 
however, have to be very large in order to have significant
effects. In our case, however, we do not need them.  Thus, assuming
that they are at their natural size, they are not expected to
significantly affect the important ingredients of our model: The
number of maxima, their widths and their rough locations.

So far we considered only the zero modes. Heavy modes, namely, the
Kaluza-Klein tower of the SM fields and heavy SM singlets fields,
cannot be ignored since they can significantly contribute to SM rare
processes. In particular, proton decay, $N - \bar N$ oscillation,
meson mixing and lepton number violating processes.  The most severe
bound on the properties of the heavy fields is usually obtained form
proton decay data.  In models where the fundamental scale is low, very
small overlap between the leptons and quarks wave functions was
proposed as a way to avoid rapid proton decay \cite{AS}. The required
tiny overlap can be achieved in models with Gaussian wave
functions. However, as we explained, such models require significant
fine tuning. Thus, proton stability can be explained by having
a very high fundamental scale or by adding new ingredients to the
model.  On the other hand, the data on other rare processes can be
accommodated in models with a relatively low fundamental scale (see
e.g. \cite{KT}).

To conclude we summarize our main results.  We showed that realistic
models of split fermions can be constructed using two scalars in a
world with one compact extra dimension compactified on an orbifold. We
found that models of split fermions with Gaussian overlaps require
large 5d Yukawa couplings, and are therefore unnatural.  We construct a
realistic two scalar model where the overlap is not purely Gaussian
which accounts naturally for the quark flavor parameters.

\acknowledgments
It is a pleasure to acknowledge helpful discussions with David
E. Kaplan and Martin Schmaltz.  Y.G. thanks the Aspen Center for
Physics for hospitality while this work was started.  Y.G. is
supported in part by the Israel Science Foundation (ISF) under Grant
No.~237/01, by the United States--Israel Binational Science
Foundation (BSF) through Grant No. 2000133, and by the fund for the
promotion of research at the Technion.

\appendix

\section{The GGH model} \label{ap_GGH}
Here we analyze the GGH mechanism of fermion separation \cite{GGH}.
We show that in this model it is rather unnatural to obtain sharp
localization and separation of the fermions at the same time.

With our scaling [see eqs. (\ref{scale1}) and (\ref{scale2})], 
the GGH model is described by 
\bea{\cal S}_\psi&=&\int dx_\mu\int_0^1 du
\left\{\bar \psi\left[i\gamma^\mu\partial_\mu-
{1\over L}\gamma_5\partial_5-{ fa\over L}
\left(1-{\kappa\over L^2 M_*^2}\left(L^2
\partial_\mu\partial^\mu-\partial_5^2
\right)
\right)\varphi\right]\psi\right.\nonumber\\
&+&\left.
{\tilde v^2\over L}\left[
L^2 {1\over2}\partial^\mu\varphi\partial_\mu\varphi_i
-{1\over2}\partial_5\varphi\partial_5\varphi
-{a^2\over2}\left(\varphi^2-1\right)^2\right]\right\} 
\,. \label{L5dGGH}
\eea
When $a$ is large, $g(u)$ is approximated by
\bea
g(u)\approx fa\tanh (au)\left[1
-{2\kappa\over b^2}{\rm sech}^2(au)
\right] {\rm ~~~for~~~} u\lsim 1/2
\label{psi5'ap}
\,,
\eea
with $b\equiv {L M_*/ a}\gg1$. The fermion wave function in the
fifth dimension is given by [see eq. (\ref{solkx})]
\bea
y(u\lsim1/2)\approx N\exp\left\{-f \left[
\ln\left[\cosh\left( a u\right)\right]+{\kappa\over ab^2}
{\rm sech}^2\left( a u\right) \right]\right\}\,.
\label{psi5'an}
\eea
The maximum is found at
\bea\label{um'}
\umax={1\over a}{\rm arctanh}\sqrt{1-{b^2\over 2\kappa}}
\,,
\eea
and the width is given approximately by
\bea\label{WGGH}
\Gamma\approx 
{1\over a \sqrt{ f}}\cdot {1\over \tanh (a \umax)}
\,.
\eea
Note few problematic points:
\begin{itemize}
\item
In order to have a maximum in the bulk, very large coupling is needed,
$\kappa >b^2/2$ \cite{GGH}.
\item
When $\kappa \gg b^2/2$, the peak of the wave function is far from the
origin. In that case, however,
$\umax\propto\log(\kappa)$. Consequently, to separate the fermions in
the extra dimension, different flavors must have large hierarchy in
their corresponding values of $\kappa$.
\item
When $\umax$ is small the width is proportional to $1/\umax$.
Consequently, a very large $f$ is needed in order to get a sharply
localized wave functions.
\end{itemize}


\newpage
\vspace*{-1.5cm}


\hspace*{-2.8cm}
\raisebox{0.cm}{\includegraphics[width=5.7cm,height=7cm]{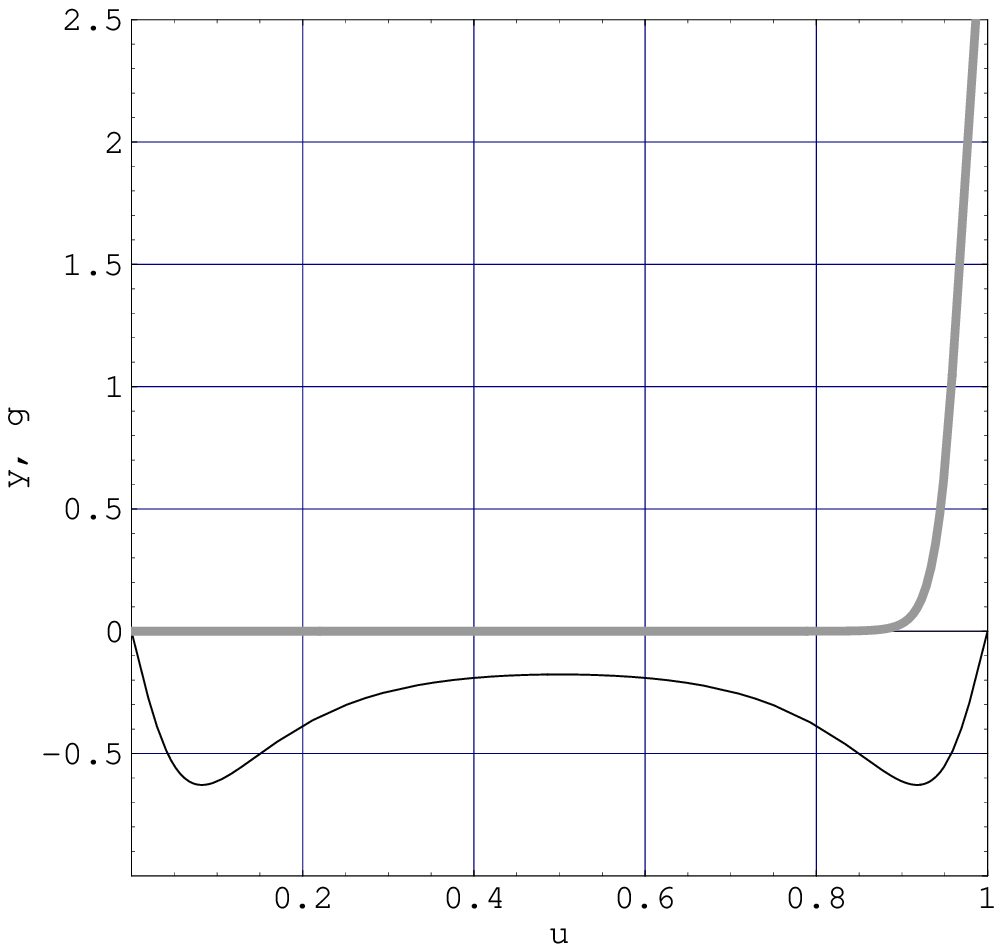}}
\hspace*{.3cm}
\raisebox{0.cm}{\includegraphics[width=5.7cm,height=7cm]{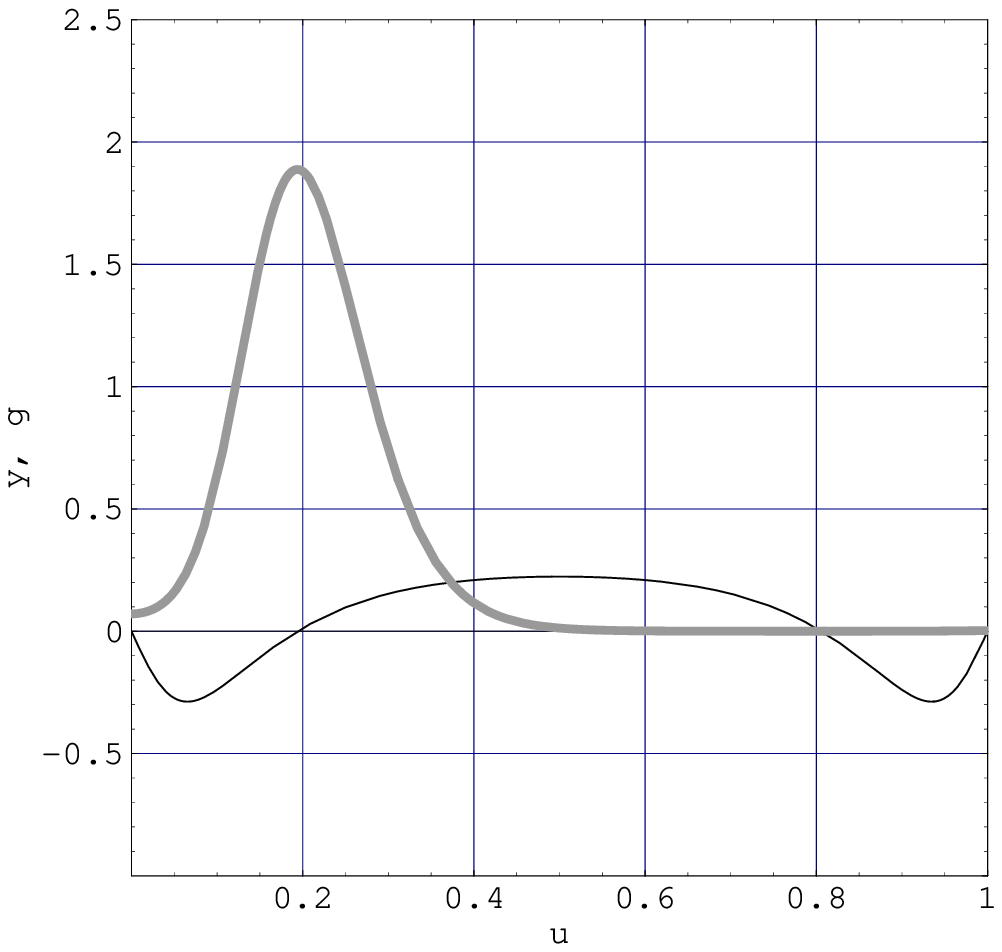}}
\hspace*{.3cm}
\raisebox{0.cm}{\includegraphics[width=5.7cm,height=7cm]{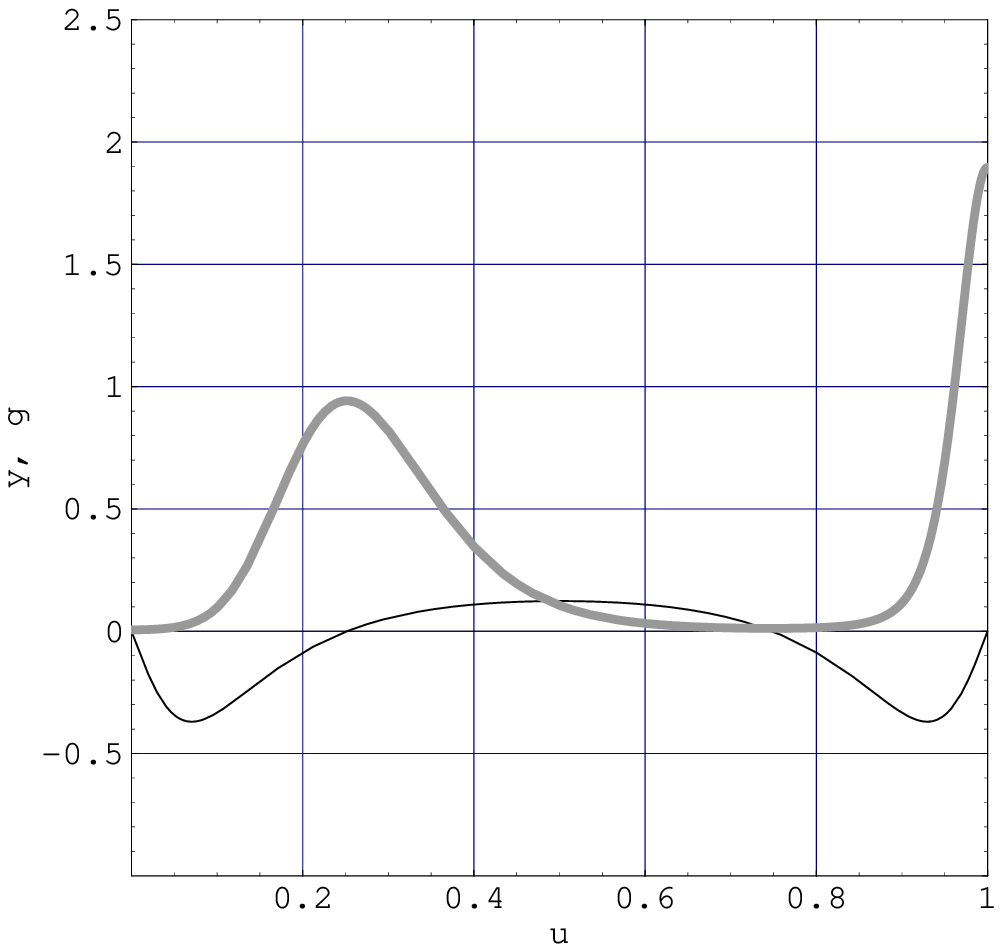}}
\vspace*{-.5cm}

{\small \hspace*{0.0cm} {\bf (1a)}\hspace*{5.35cm} {\bf (1b)}\hspace*{5.35cm} {\bf (1c)}}\\
\vspace*{-.4cm}

\hspace*{-.6cm} {\small Fig. 1: The typical shapes of the scalar vevs and
  the fermion wave functions.
The thin black curves correspond to the
  effective scalars vev, $g(u)$ [eq. (\ref{gu})]. 
The thick gray curves correspond to the
  fermion wave functions, $y(u)$ [eq. (\ref{CM})].
The relevant parameters are $a_1=5$, $a_2=17$, and $X=1.15,0.75,0.85$
for figs. {\bf (1a), (1b)} and {\bf (1c)} respectively.
In this example $X_b\sim0.87$ [eq. (\ref{twoM})].}

\vspace*{1cm}
 
\hspace*{-2.85cm}
\raisebox{0.cm}{\includegraphics[width=9.4cm,height=7cm]{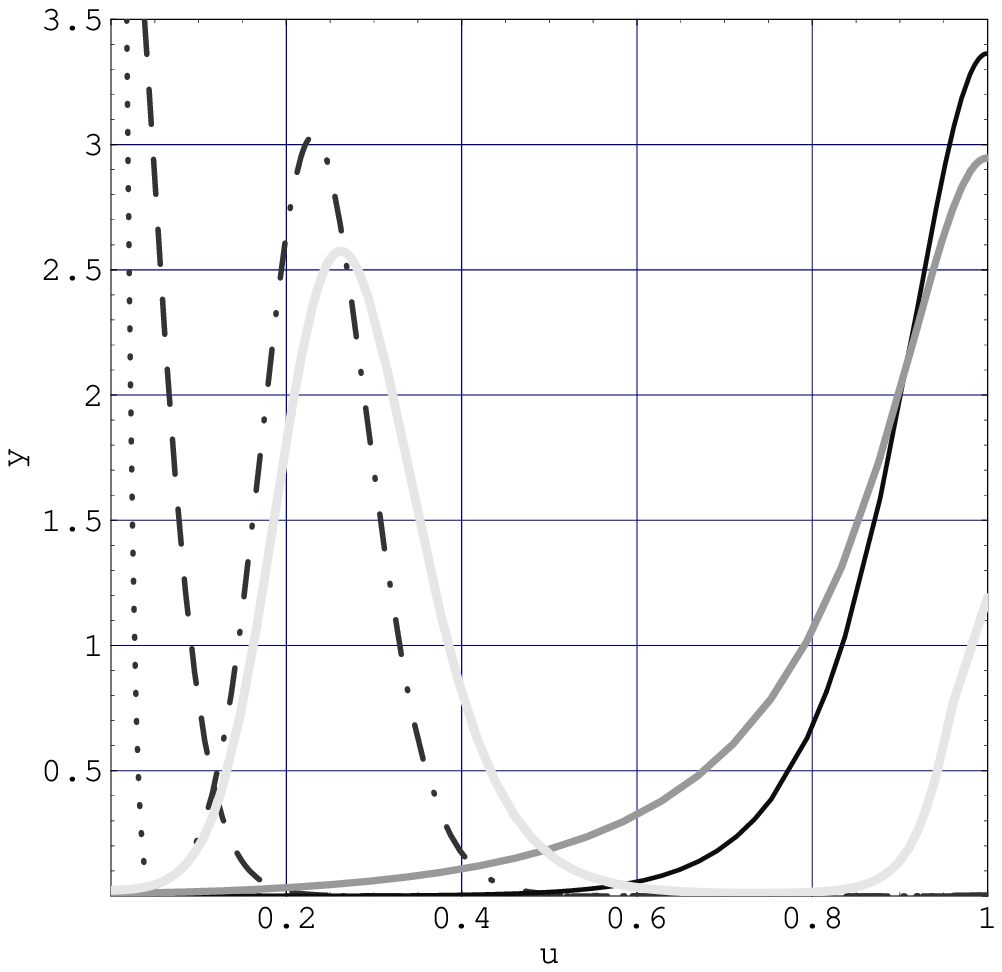}}
\hspace*{.2cm}
\raisebox{0.cm}{\includegraphics[width=9.4cm,height=7cm]{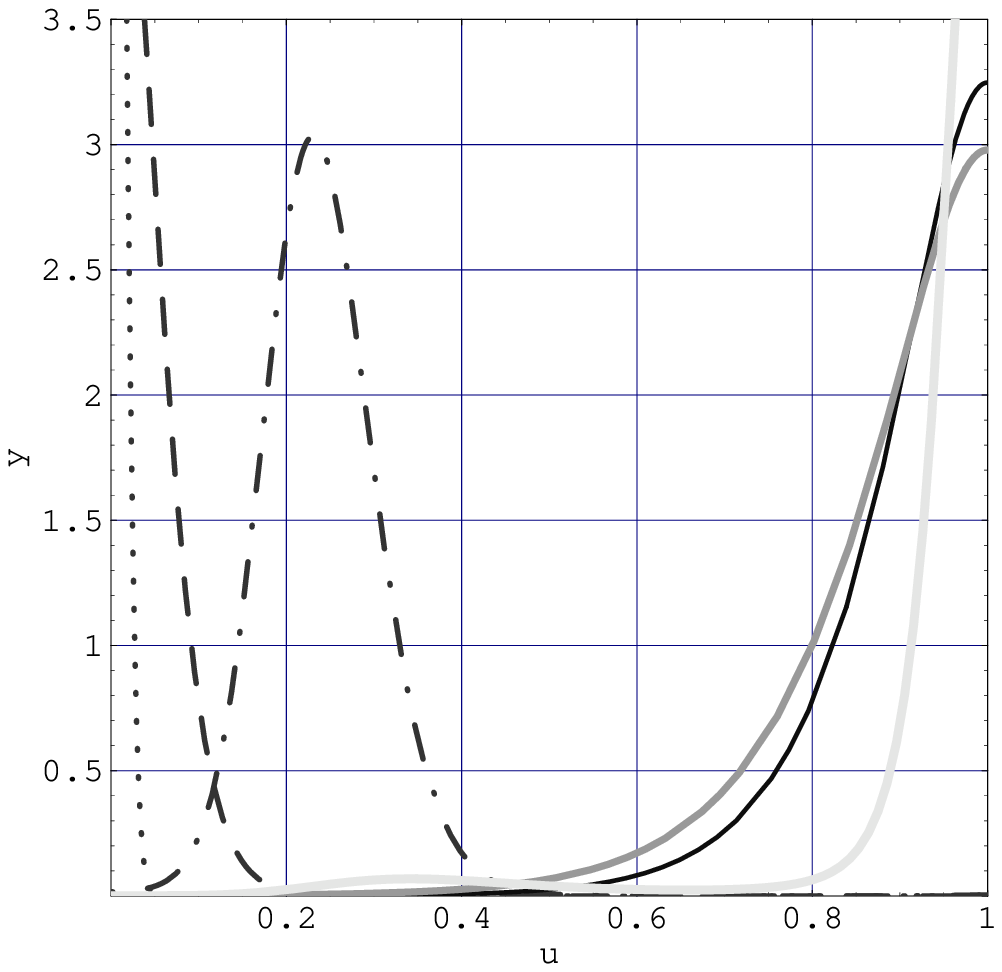}}
\vspace*{-1.cm}

{\small \hspace*{2.2cm} {\bf (2a)}\hspace*{9.05cm} {\bf (2b)}}\\
\vspace*{-.4cm}

\hspace*{-.6cm} {\small Fig. 2: 
The configuration of the quark wave functions in the
extra dimension.
Figs {\bf (2a), (2b)} are related to overlaps between
 the up and down type quarks respectively. 
$y^{Q_{1,2,3}}$ are
 the dotted, dashed, and dashed-dotted curves
 respectively (shown in both figs). $y^{u_{1,2,3}}$ and $y^{d_{1,2,3}}$
 are the thin black, mid gray and thick
 light curves of figs. {\bf (2a)} and  {\bf (2b)}
respectively. The relevant parameters are given in [eq. (\ref{MSbar})].}

\end{document}